\documentclass[%
 reprint,
 amsmath,amssymb,
 aps,
]{revtex4-1}

\usepackage{enumitem}
\usepackage{color}

\usepackage{graphicx}
\usepackage{dcolumn}
\usepackage{bm}

\begin{document}


\title{Off-axis runaway-electron seed formation, growth and suppression}

\author{L. F. Delgado-Aparicio$^{1}$, D. Del-Castillo-Negrete$^{2}$, N. C. Hurst$^{3}$, P. VanMeter$^{3}$, M. Yang$^{2}$, J. Wallace$^{3}$, A. F. Almagri$^{3}$, B. E. Chapman$^{3}$, K. J. McCollam$^{3}$, N. Pablant$^{1}$, K. Hill$^{1}$, M. Bitter$^{1}$,  J. S. Sarff$^{3}$ and C. B. Forest$^{3}$}

\affiliation{$^{1}$Princeton University Plasma Physics Laboratory (PPPL), Princeton, NJ, 08540, USA \\ 
                  $^{2}$Oak Ridge National Laboratory (ORNL), Oak Ridge, TN, 37831, USA \\ $^{3}$University of Wisconsin-Madison, Madison, WI, 53706, USA}

\date{\today}

\begin{abstract}
Novel x-ray detection technology enabled the first profile measurements of the birth and growth dynamics of runaway electrons (REs) at the edge of tokamaks during quiescent RE studies at the Madison Symmetric Torus. The formation of an off-axis 
RE seed with linear growth rates has been resolved for 
low energies, a hollow streaming parameter and large electric fields ($E_{\parallel}/E_{D}$) in agreement with theory and simulations. Secondary exponential growth rates have also been spatially resolved for the first time and are consistent with a convective transport of the order of the Ware pinch and energies up to $10^3\times T_{e,0}$. Numerical simulations are shown to reproduce the experimental observations including the off-axis runaway electron generation, radial transport and exponential growth at the core, as well as suppression due to $m=3$ resonant magnetic perturbations.
 
\end{abstract}

\maketitle


The intrigue and fascination with the physics of runaway electrons (REs) is a century old. REs typically experience a ``\textit{free-fall}'' acceleration which exceed the collisional drag force exerted by elastic, ionization and Coulomb interactions; this frictionless feature allows them to achieve high velocities close to the speed of light and energies up to few tens of MeV. This phenomenon was first pointed out conceptually by C. T. R. Wilson \cite{ctrW24}, A. S. Eddington \cite{asE26} and R. G. Giovanelli \cite{fgG46} when describing the acceleration of $\beta$-particles in the strong electric fields of thunderclouds and chromospheric flares. Since then, RE phenomena have been related to various astrophysical occurrences \cite{jRD06}, Rayleigh-Taylor driven plasma jets in small experiments \cite{rsM18}, excitation of Alfv$\acute{e}$n waves \cite{lC21}, reduction in fusion yield during dense plasma focus experiments \cite{ejL14} among many others. 
REs are also one of the main ``\textit{Achilles heels}'' in the elusive goal of controlled nuclear fusion for energy production. In particular, electric fields generated during tokamak start-up and disruption can generate high energy RE which, if not controlled, can cause serious damage to plasma-facing components of fusion reactors \cite{vS09}. 
Thus, understanding the genesis of the seed, growth and mitigation is essential to avoid the formation of RE currents in SPARC, ITER and beyond. 

The majority of ``startup'' (S), ``quiescent'' (Q) and ``disruption'' (D) RE studies conducted over the last four decades dealt with core ``keV'' temperatures featuring non-thermal particle distributions and photon-emission energies up to few tens of MeV; in-short, ratios of $E_{photon}/T_{e,0}\sim$1000-30000. Several review papers \cite{rsG14,bnB19} revealed important properties of 20-40 MeV REs when emitting synchrotron radiation \cite{cPZ14,raT19}; spectrometers, single diode measurements and pin-hole imagers 
focused also on gamma-ray energies between 2-20 MeV \cite{cPZ17,cPZ18}. It is well known however that the energy spectrum of emitted photons in a Bremsstrahlung encounter has 
a peak at about 1/4 of that of its incident energy \cite{lfDA10,smS85} suggesting therefore that most radiation measurements were done with electrons at ``\textit{terminal}'' velocities near the speed of light. 
Early detection and understanding of low energy phenomena is thus critical for low-risk control of REs preventing the need of  dissipation strategies at dangerously high energies \cite{ahB17}.

A significant payback for developing active control strategies could be achieved focusing on the physics and signatures of the low energy seed formation. However, 
the typical suite of diagnostics used in tokamaks is not tailored to low-energies since soft x-ray (SXR) systems often suffer from 
saturation while hard x-ray (HXR) and $\gamma$-detectors often hit a finite noise floor. 
Therefore, the wide energy range expected from the seed condition to its full energy cannot be surveyed by any single diagnostic.

In this Letter we present novel results obtained with a unique and versatile multi-energy soft x-ray (ME-SXR) pinhole camera 
detecting the seed profile and also measuring the evolution of the RE-growth with an unprecedented level of detail. When specified, an entire row or column of pixels could also be effectively used for a different energy range obtaining coarse spectral resolution without losing spatial information. The ability to set a threshold energy at an arbitrary value with constant resolution is a significant improvement over metallic foil systems 
\cite{lfDA16}-\cite{lfDA21}. The ME-SXR circumvents thus the limitations of conventional x-ray systems providing unprecedented technological improvement in throughput (up to $\sim3\times10^{10}$ photons/cm$^2$/s) with extremely low-noise and high signal-to-noise-ratio (SNR), enabling early-detection, imaging 
and simultaneous discrimination at seed-energies of the order of $\sim$(20-200)$\times T_{e,0}$. 
\begin{figure}[t!]
\includegraphics[scale = 0.9]{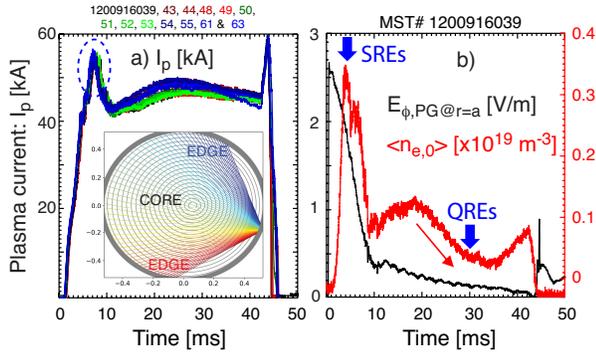}
\caption{(Color online) a) Reproducible traces of $I_{p}$ with the geometry of the ME-SXR pin-hole camera, and b) an overlay of the toroidal electric field ($E_{\phi}$) and core line-integrated density ($\langle n_{e,0}\rangle$) during the runaway electron experiments.} \label{fig:Fig1}
\end{figure}

The primary early-acceleration is one of the simplest kinetic properties of plasmas since it involves a positive net force between the accelerating electric field ($E_{\parallel}$) and the background drag such that, $eE_{\parallel}>m_{e}\nu_{e}(v_{e})v_{e}$, where $\nu_{e}(v_{e})=n_{e}v_{e}\sigma_{C}\propto v_{e}^{-3}$ is the collision frequency and $\sigma_{C}=e^4\ln\Lambda/4\pi\epsilon_{0}^2m_{e}^2v_{e}^{4}$ is the Coulomb cross-section. That is, if the electric field is weak, only high-energy electrons will experience runaway acceleration, but if the field is large and comparable with the Dreicer field ($E_{D}\equiv n_{e}e^{3}\ln\Lambda/4\pi\epsilon_{0}^2T_e$) the runaway condition may be satisfied even for the thermal bulk of the distribution function. Conceptually, this seed electrons leave behind a deficit in the distribution which is filled in by diffusion in velocity-space allowing new electrons to accelerate. The role of a more efficient exponential secondary generation occurs when REs amplify the low-energy preexisting seed. 
\begin{figure}[t!]
\includegraphics[scale = 0.7]{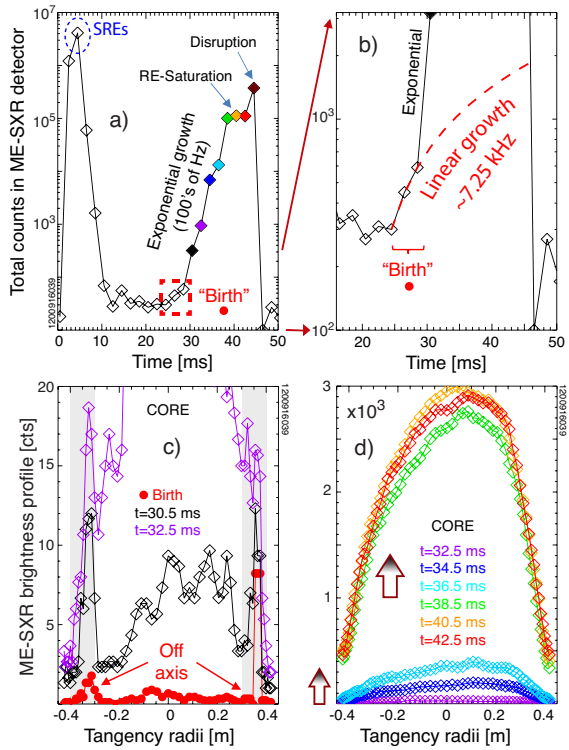}
\caption{(Color online) The time-history of the total number of counts in frame is depicted in -a) while the initial linear growth and its off-axis profile is shown in -b) and -c); the growth for the secondary exponential phase is shown in d).} \label{fig:Fig2}
\end{figure}

The experiments reported in this Letter were conducted at the Madison Symmetric Torus (MST) that can produce REs in steady tokamak scenarios \cite{sM20} with low current and toroidal magnetic field obtaining plasmas with core temperatures and densities of the order of 0.1 keV and $<0.1\times10^{19}$ m$^{-3}$. Reproducible tokamak discharges with $I_{p}=40-50$ kA and $B_{\phi,wall}$=1.4 kG are routinely obtained as shown in Fig. 1-a). 
Density thresholds for both RE onset and suppression are determined with simple variations in gas puffing [see Fig. 1-b)], and the presence of runaways is usually detected via emission of non-thermal x-ray photons (see Fig. 2).

For the early startup (SRE) and quiescent (QRE) RE-experiments the entire detector was operated at a threshold-energy $\gtrsim$2 keV to increase the SNR at low-densities; the core thermal plasma has negligible x-rays above 1 keV. The time-history of the total counts in the SXR detector is depicted in Fig. 2-a) and shows a clear non-thermal maximum at the startup (SRE at $t\sim4-5$ ms) where ``hard'' x-rays are expected due to the high electric field needed for the ionization,  and later during the growth and peak of the QRE experiment; the ME-SXR demonstrates meaningful sampling at a high dynamic range over six orders of magnitude with small floor and no saturation. 
\begin{figure}[t!]
\includegraphics[scale = 1.0]{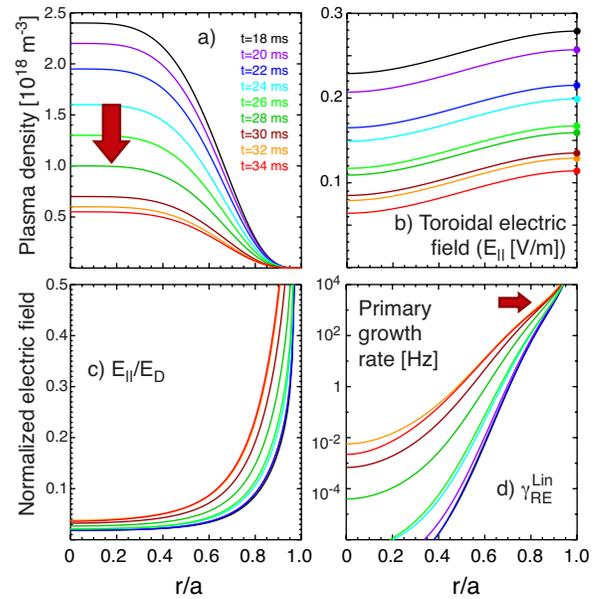}
\caption{(Color online) Spatially-resolved time-history profiles for a) plasma density, b) toroidal electric field, c) the ratio of $E_{\parallel}/E_{D}$, and the linear growth [see Eqn. (2)].} \label{fig:Fig3}
\end{figure}
\begin{figure*}[t!]
\includegraphics[scale = 0.9]{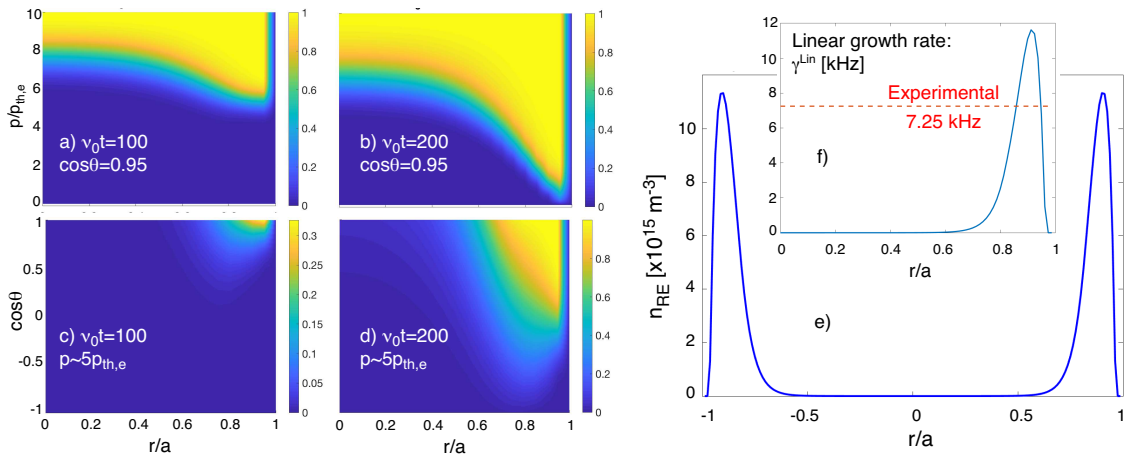}
\caption{(Color online) 
Time evolution of 2D cuts of the 3D probability of runaway function, $P_{RE}(p,\theta,r,t)$, at constant pitch angle in panels (a)-(b), and at constant momentum in (c)-(d). Panel (e) shows the RE density $n_{RE}(r,t)$ and (f) the linear growth rate $\gamma \sim  dn_{RE}/dt$  for the MST experiments of interest where $\nu_0$ is the thermal collision frequency.} \label{fig:Fig4}
\end{figure*}

The QRE study consisted in a pre-programmed density ramp-down [$\langle n_{e}\rangle = 0.16\to0.05\times10^{19}$ m$^{-3}$ from 18 to 35 ms as shown in Fig 1-b)] generating fast electrons as $E_{D}\propto n_{e}$. Using the total counts in frame as shown in Fig. 2-a) and -b) we clearly identify a first birth signature with a primary linear growth of approximately 7 kHz. A ``\textit{zoom}'' on the linear-phase is depicted in Fig. 2-b) and shows how this primary growth cannot reproduce the experimental data when extrapolated at later times since a secondary exponential mechanism must be at play.  The spatial profiles during the linear growth shown in Fig. 2-c) suggest that the seed forms first at an off-axis location ($r/a\sim0.8$, $r_{T}\sim0.4$ m) far from the core; these off-axis signatures are also observed in the early phase of the exponential growth. 
This second phase amplifies the pre-existing seed with a growth rate of $\sim500-700$ Hz.
 
The emergence of the seed population in the plasma periphery instead of that in at the magnetic axis is qualitatively and quantitatively consistent with a region with lower density and thus reduced Dreicer fields. The data shown in Figs. 3-a) and -b) describe the time evolution of the main plasma profiles during the QRE study. The values of the toroidal electric field ($E_{\parallel}$) were derived using equilibrium reconstructions \cite{cF94,jA04} and Faraday's equation constrained by internal loop-voltage measurements at the wall 
 \cite{bE96}. As the density and Dreicer fields remain peaked but decreasing in time, the values of the $E_{\parallel}/E_{D}$ ratio remain hollow reaching values of 2-5\% at the core but $\sim10-20\%$ at $r/a\sim0.8$ and even higher at the periphery (40-50\%). This is consistent with finding large streaming parameters $\xi(r)\approx E_{\parallel}(r)/E_{D}(r)$ \cite{hK79,pcdV19} in comparison of those found by Spitzer and Harm for thermal plasmas ($\xi\ll1$). Large values of $E_{\parallel}/E_{D}$ suggest therefore that a significant fraction of the outward thermal plasma is \textit{running away}. 


The experimental growth rate of the primary Dreicer mechanism is linear and can be approximated using the density of the \textit{Maxwellian}-tail with $v_{e}/v_{th,e}\gtrsim\sqrt{E_{D}/E_{\parallel}}$ and a timescale comparable to one collision time as $\gamma^{Lin}_{RE} \sim \frac{eE_{\parallel}}{m_{e}v_{th,e}} \exp\left( -\frac{E_{D}}{2E_{\parallel}} \right)$. 
For fields of the order of 9-16\% of Dreicer's, the seed should occur at values of (3-4)$\times v_{th,e}$ or electrons which are (27-64)$\times$ less collisional than thermals. 
A plot of these growth rates is depicted in Fig. 3-d) 
and show its strong hollow nature matching the experimental linear growth rate of few kHz at the periphery instead of that the magnetic axis ($\ll$1 Hz). The formalism for the linear growth rate was further refined in the non-relativistic limit ($E_{CH}\ll E_{\parallel}\ll E_{D}$; $E_{CH}=E_{D}\cdot T_{e}/m_{e}c^2$ is the Connor-Hastie field) using a rigorous treatment of kinetic theory 
by Gurevich, Kruskal and Cohen \cite{avG61}-\cite{rhC76} but gives very similar results.

To compare these experimental findings with theory we consider a time-dependent 3D Fokker-Planck (FP) model for the electron distribution function, $f(p,\theta,r,t)$, where $p$ is the momentum, $\theta$ the pitch angle, and $r$ the minor radius of the tokamak. The model includes the standard physics of electric field acceleration, radiation damping, Coulomb drag and momentum and pitch angle scattering, in addition to radial diffusion. 
However, contrary to the common used approaches based on the direct numerical solution of the FP equation we use the Backward Monte Carlo (BMC) method that uses the Feynman-Kac theory to compute the probability density function, $P_{RE}(p,\theta,r,t)$, that an electron with coordinates $(p,\theta,r)$ runs away at a time $\leq t$. Once $P_{RE}$ is known, the density of produced RE is given by $n_{RE}(r,t)=\int_\Omega P_{RE} f_{M} d\Omega$ where $f_M(p,\theta)$ is the equilibrium initial Maxwellian distribution and $d\Omega$ is the phase space volume element. Further details on the BMC method and the physics model can be found in \cite{gZ17}-\cite{dDCNB21}. 
In the calculations presented here, the spatiotemporal evolution of the electric field, collisional frequencies and diffusivity are inferred from the experimental data including the profiles in Fig. 3. Consistent with the experimental findings, the BMC calculations depicted in Fig. 4 show large values of $P_{RE}$ strongly localized in the edge that result on an off-axis RE density profile. Most importantly, the mean of the computed growth rate profile is consistent with theory and the experimental value measured at MST. 
\begin{figure}[t!]
\includegraphics[scale = 0.8]{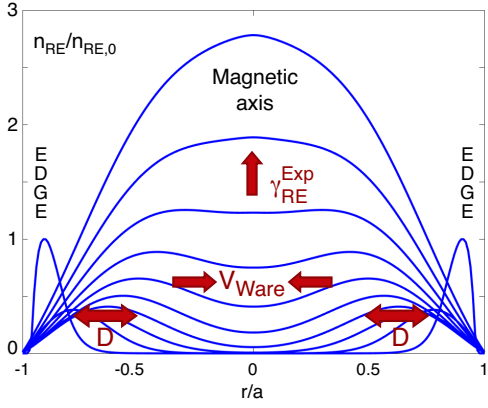}
\caption{(Color online) Solution of the convection-diffusion transport of REs with a transient edge seed, diffusion, Ware pinch and a non-linear  exponential growth near the axis.}\label{fig:Fig5}
\end{figure}
\begin{figure}[t!]
\includegraphics[scale = 0.8]{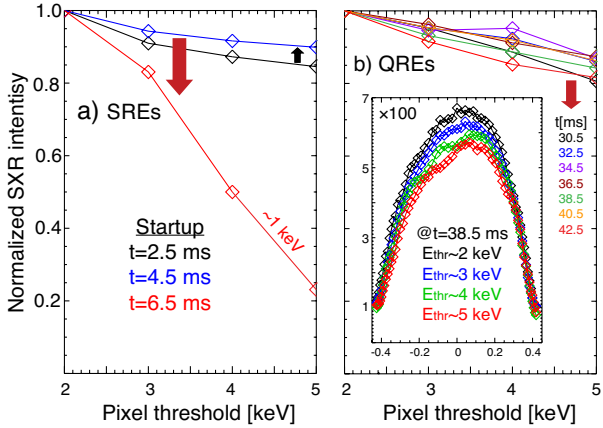}
\caption{(Color online) Experimental normalized x-ray intensities $\geq$2 keV during the a) SRE and b) QRE phases.} \label{fig:Fig6}
\end{figure}

As shown in Figs. 2-a) and -d), the second phase of the RE evolution is characterized by an exponential growth of the fast-electron density at the plasma core. This density peaking results from the combined effect of the Drecier production of primary RE at the edge (as described above), radial transport, and the ideal avalanche production of secondary REs \cite{mR97}.
The RE transport consists of radial diffusion and the Ware pinch, which, as discussed in \cite{eN15,cMcD19}, can transport trapped RE towards the core where they can be de-trapped and further accelerated. To capture this physics we considered the following simple transport model for the RE density evolution, 
$\partial_t n_{RE}=-\nabla \cdot \Gamma + \gamma_{RE} n_{RE}$, where 
$\Gamma= - D \partial_r n_{RE} + V_{Ware} n_{RE}$ with $V_{Ware}=- \alpha E_{\parallel}/B_\theta$ the Ware pinch, $D$ the spatial diffusivity and $\gamma_{RE}$ the avalanche growth rate. The low-temperature and impurity content from MST plasmas provide large pitch-angle scattering due to collisions with partially ionized impurities (e.g. screening effects must be included), as well as an enhanced diffusive and convective transport \cite{cMcD19}. Figure 5 shows the short-time spatiotemporal evolution of $n_{RE}$ according to this transport model using the seed RE density in Fig. 4-e) as initial condition. The long-time dynamics is dominated by the avalanche source $\gamma_{RE} n_{RE}$ which consistent with the experiment leads to the exponential peaking of the density at the magnetic axis. The enhanced experimental value of $\gamma_{RE}$ over the ideal approximation \cite{mR97} reflects the need to include the effect of partial screening of bound electrons \cite{lH19,lH192}. In particular, for the experiments under consideration, MST's aluminum wall might seed the core plasma with medium-Z ($Z_{Al}\sim3\to8$) impurities, in addition to the expected carbon ($Z_{C}\sim4$) and oxygen ($Z_{O}\sim4\to6$) background for temperatures in the range of $T_{e}\in[20-100]$ eV. 
\vspace{-0.5mm} The continuous energy gain during the exponential growth can be measured through the energy-dependence of x-ray emission. The detector configuration allows to resolve the time-history profiles in four energy ranges with thresholds at 2, 3, 4 and 5 keV demonstrating an innovative capability never tested with REs. The strong dependance with energy 
shows that a ``hot'' component could be easily identifiable when the intensity ratios approach unity (see Fig. 6). 
For the initial SRE phase, the ``hot'' component (at $10^2-10^3\times T_{e,0}$) increases energy from 2.5 to 4.5 ms after the plasma initiation but disappears quickly due to a dynamic interplay between lowering the electric field and raising the density. For the QREs phase the intensity ratio $I_{5keV}/I_{2keV}$ typically grows from 0.8$\to$0.9 suggesting the presence of an even ``hotter'' component up $\gtrsim10^3\times T_{e,0}$. An example of the peaked radial profiles with a slight asymmetry possibly due to the Shafranov-shift compressing flux-surfaces at the LFS is shown in the see inset in Fig. 6-b).
\begin{figure}[t!]
\includegraphics[scale = 0.75]{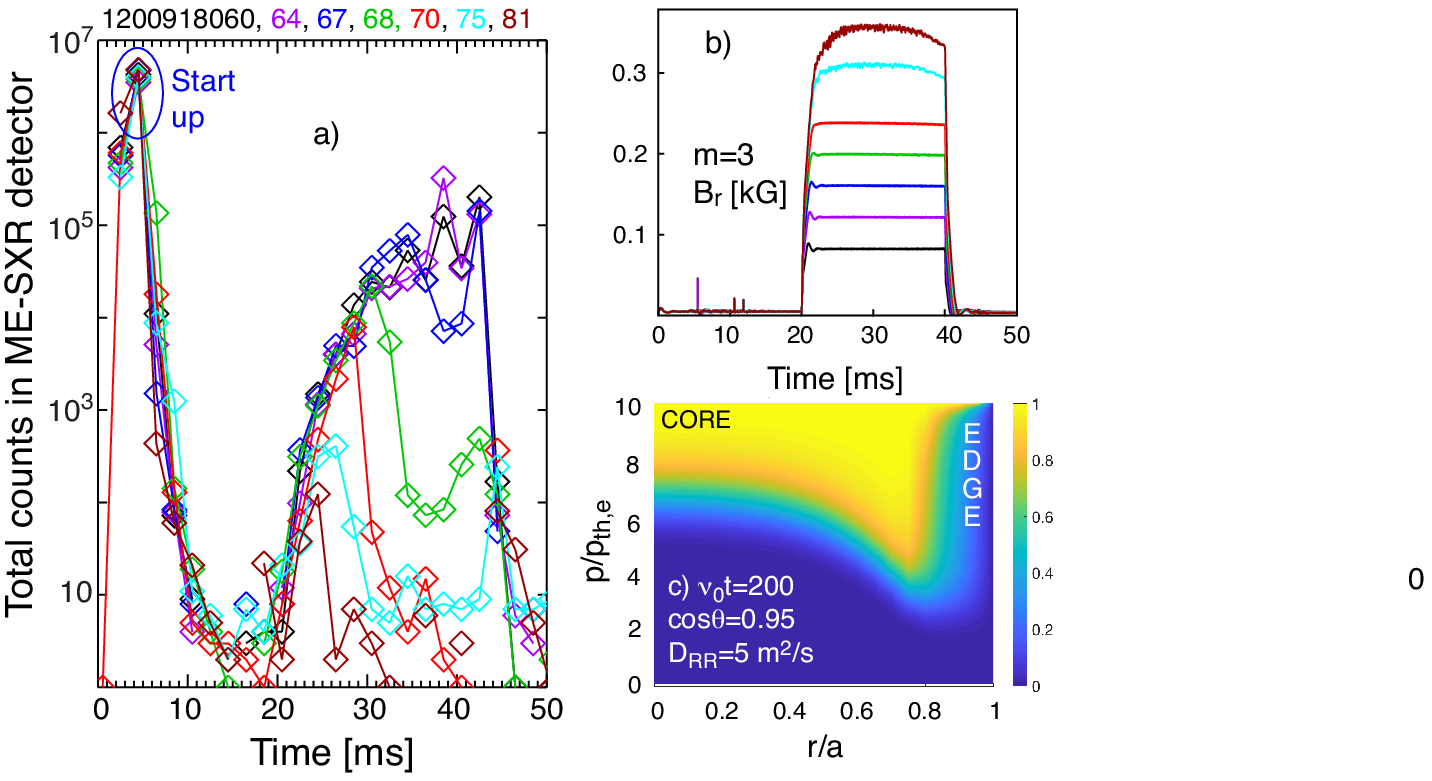}
\caption{(Color online) Scan of $B^{m=3}_{r}$ intensity suppressing the exponential growth in MST. The BMC probability of finding REs for $D_{RR}\sim$ 5m$^{2}/s$ is shown in c).} \label{fig:Fig7}
\end{figure}

\vspace{-0.5mm} Last but not least, resonant magnetic perturbations (RMPs) can be used to introduce stochasticity in the plasma edge - where the seed originate - increasing electron transport and minimizing 
the transfer of energy that REs can achieve. 
Data in Fig. 7 depicts a detailed shot-to-shot scan of $m=3$ RMP 
amplitude and its effectiveness suppressing RE seed at the edge. As discussed by Munaretto \cite{sM20} and recently by Cornille \cite{bC20} using NIMROD, the $m=1$ RMP will have little to no effect on the topology while the $m=3$ RMP produces a broad region of stochasticity at the edge ($r/a>0.7$) which allows for rapid loss of REs. To model these experimental results we use the BMC method with a diffusivity $D=D_{RR} F(r)$ where $D_{RR}=\pi q v_{\parallel} R (\delta B/B)^2=5 {\rm m}^2/{\rm sec}$ is the Rechester-Rosenbluth diffusivity and $F(r)=\left\{1+\tanh \left[ \left(r-r_D\right)/L_D\right] \right\}$ is a simple model of the penetration of the RMP with $r_D=0.85 \,a$ and $L_D=0.02 \,a$. Consistent with the experimental findings, the numerical simulations show a significant reduction of the probability of runaway, $P_{RE}$, for $r/a\geq 0.8$, compared with the results in the absence of RMP shown in Fig. 4-b). 

In summary, 
the formation of an off-axis RE seed with a linear growth has been resolved for photon energies $\sim$(20-200)$\times T_{e,0}$ using a novel multi-energy SXR camera. The emergence of the seed population in the plasma periphery instead of that at the core is consistent qualitatively and quantitatively with a lower electron-density and Dreicer fields as well as hollow $E_{\parallel}/E_{D}$ ratios and streaming parameters, all in agreement with theory and numerical simulations. Secondary exponential growth rates have also been spatially resolved for the first time and are consistent with an enhanced diffusivity, convective transport of the order of the Ware pinch and energies up to $10^3\times T_{e,0}$. State of the art calculations with the BMC code successfully resolved the fully space- and time-dependent dynamic scenarios including that of RMP suppression at the edge even in regions with large $E_{\parallel}/E_{D}$.

This work was supported by the U.S. DOE-OFES, under Contract No. DE-AC02-09CH11466, and L.F.D-A's 2015 DOE Early Career Award Research Program. D.d-C-N. and M. Yang were sponsored by the U.S. DOE-OFES, Office of Advanced Scientific Computing Research, SciDAC program, at the Oak Ridge National Laboratory, operated by UT-Battelle, LLC, for DOE under Contract DE-AC05-00OR22725. The experiments were conducted at the Wisconsin Plasma Physics Laboratory (WiPPL), a research facility supported by the DOE-OFES, under Contract No. DE- SC0018266, with major facility instrumentation developed with the support from the NSF, under Award No. PHY 0923258. The MST Tokamak Thermal Quench Program is funded under DOE-OFES, Award No. DE-SC0020245.

\end{document}